\documentclass[sigconf]{acmart}

\usepackage{amsmath}
\usepackage{amsthm}
\usepackage{cancel}
\usepackage{mathtools}
\usepackage{graphicx}
\usepackage{subfigure}
\usepackage[table]{xcolor}
\usepackage{xcolor}
\usepackage{siunitx}
\usepackage[inline]{enumitem}
\usepackage{geometry}
\usepackage{booktabs}
\usepackage{multirow}
\usepackage{acronym}
\usepackage[skip=2pt]{caption}
\usepackage{bm}
\usepackage{booktabs}
\usepackage{tabularx}
\AtBeginDocument{%
  }

\DeclareMathOperator*{\argmin}{argmin} 

\newcommand{\code}[1]{{\ttfamily#1}}

\newcommand{\headernodot}[1]{\noindent\textbf{#1}}
\newcommand{\header}[1]{\headernodot{#1.}}

\acrodef{ctg}[CTG]{controllable text generation}
\acrodef{mips}[MIPS]{maximum inner product search}
\acrodef{gr}[GR]{generative retrieval}
\acrodef{bpb}[bpb]{bits per byte}
\acrodef{bpt}[bpb]{bits per token}
\acrodef{llm}[LLM]{large language model}
\acrodef{docid}[docID]{document identifier}
\acrodef{kl}[KL]{Kullback–Leibler}
\acrodef{ndcg}[nDCG]{Normalized Discounted Cumulative Gain}
\acrodef{mrr}[MRR]{Mean Reciprocal Rank}

\acrodef{diffugr}[DiffuGR]{diffusion generative retrieval}

\AtBeginDocument{%
  }

\setcopyright{acmlicensed}
\copyrightyear{2026}
\acmYear{2026}
\acmDOI{XXXXXXX.XXXXXXX}

\acmISBN{978-1-4503-XXXX-X/2026/04}

\begin{document}

\title[DiffuGR: Generative Document Retrieval with
Diffusion Language Models]{DiffuGR: Generative Document Retrieval with \\
Diffusion Language Models}

\author{Xinpeng Zhao}
\email{zhaoxp1001@gmail.com}
\affiliation{%
  \institution{Shandong University}
  \state{Shandong}
  \country{China}
}

\author{Zhaochun Ren}
\authornote{Corresponding author.}
\email{z.ren@liacs.leidenuniv.nl}
\affiliation{%
  \institution{Leiden University}
  \state{Leiden}
  \country{Netherlands}
}

\author{Yukun Zhao}
\author{Zhenyang Li}
\email{yunkunzhao.sdu@gmail.com}
\email{zhenyounglee@gmail.com}
\affiliation{%
  \institution{Baidu Inc.}
  \state{Beijing}
  \country{China}
}


\author{Mengqi Zhang}
\email{mengqi.zhang@sdu.edu.cn}
\affiliation{%
  \institution{Shandong University}
  \state{Shandong}
  \country{China}
}

\author{Jun Feng}
\email{junfeng0288@gmail.com}
\affiliation{%
  \institution{Chinese Academy of Sciences}
  \state{Beijing}
  \country{China}
}

\author{Ran Chen}
\email{chenran@stu.pku.edu.cn}
\affiliation{%
  \institution{Peking University}
  \state{Beijing}
  \country{China}
}

\author{Ying Zhou}
\author{Zhumin Chen}
\email{yingzhou@sdu.edu.cn}
\email{chenzhumin@sdu.edu.cn}
\affiliation{%
  \institution{Shandong University}
  \state{Shandong}
  \country{China}
}


\author{Shuaiqiang Wang}
\author{Dawei Yin}
\email{shqiang.wang@gmail.com}
\email{yindawei@acm.org}
\affiliation{%
  \institution{Baidu Inc.}
  \state{Beijing}
  \country{China}
}


\author{Xin Xin}
\authornotemark[1]
\email{xinxin@sdu.edu.cn}
\affiliation{%
  \institution{Shandong University}
  \state{Shandong}
  \country{China}
}

\renewcommand{\shortauthors}{Zhao et al.}

\begin{abstract}
Generative retrieval (GR) reframes document retrieval as an end-to-end task of generating sequential document identifiers (DocIDs).
Existing GR methods predominantly rely on left-to-right auto-regressive decoding, which suffers from two fundamental limitations: (i) a \emph{mismatch between DocID generation and natural language generation}, whereby an incorrect DocID token generated at an early step can lead to entirely erroneous retrieval; and (ii) an \emph{inability to dynamically balance the trade-off between retrieval efficiency and accuracy}, which is crucial for practical applications.

To tackle these challenges, we propose generative document retrieval with diffusion language models, termed \emph{DiffuGR}.
DiffuGR formulates DocID generation as a discrete diffusion process. During training, DocIDs are corrupted through a stochastic masking process, and a diffusion language model is trained to recover them under a retrieval-aware objective.
For inference, DiffuGR generates DocID tokens in parallel and refines them through a controllable number of denoising steps.
Unlike auto-regressive decoding, DiffuGR introduce \emph{a novel mechanism to first generate plenty of confident DocID tokens and then refine the generation through diffusion-based denoising}. Moreover, DiffuGR also offers \emph{explicit runtime control over the quality–latency tradeoff}.
Extensive experiments on widely-applied retrieval benchmarks show that DiffuGR outperforms strong auto-regressive generative retrievers.
Additionally, we verify that DiffuGR achieves flexible control over the quality–latency trade-off via variable denoising budgets.
\end{abstract}



\received{20 February 2007}
\received[revised]{12 March 2009}
\received[accepted]{5 June 2009}

\maketitle

\section{Introduction}
\header{Generative Retrieval}
In recent years, generative retrieval (GR) has emerged as a promising paradigm that replaces traditional index-based retrieval pipelines with an end-to-end generative model capable of directly producing document identifiers (DocIDs) in response to a query \cite{DeCao2020AutoregressiveER, Tay2022TransformerMA, Wang2022ANC, Zhou2022UltronAU}.
By integrating retrieval tightly with generation
and memorizing documents with model parameters, GR offers new potential to unify retrieval and reasoning, reduce system complexity, and unlock more flexible knowledge access.

\begin{figure}
    \centering
    \includegraphics[width=0.85\linewidth]{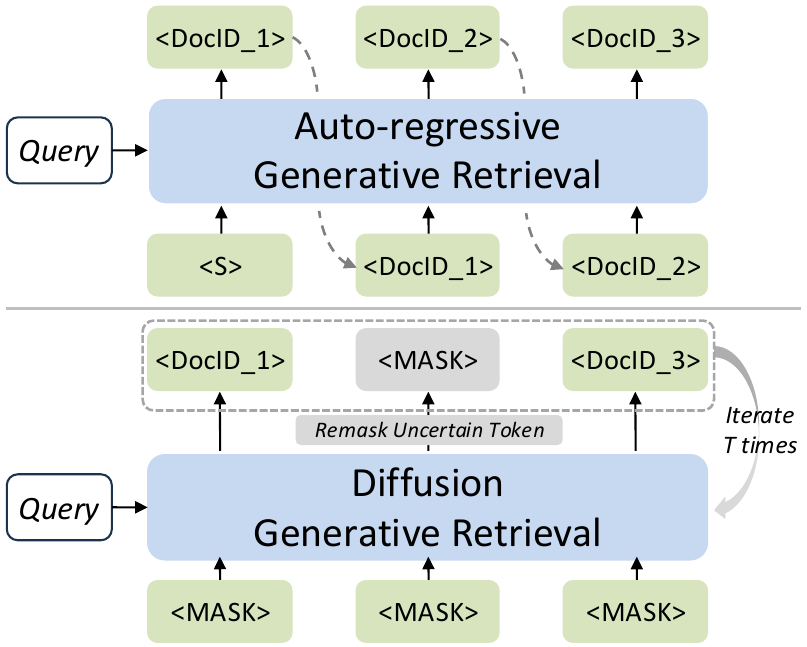}
    \caption{In contrast to auto-regressive generative retrieval (above), which generates the DocID from left to right, DiffuGR (below) efficiently generates multiple tokens in parallel and then re-masks uncertain tokens.
    The generation is progressively refined over controllable denoising steps.}
    \label{fig:ar-vs-dr}
\end{figure}

Most existing GR methods are built upon auto-regressive language models, which sequentially generate discrete identifier tokens in a left-to-right manner, as shown in the above part of Figure \ref{fig:ar-vs-dr}. 
Plenty of studies have been conducted following this research line, including constructing better DocIDs \cite{Mehta2022DSIUT,  sun2023learning, yang2023auto_search_indexer, zeng2024planning}, using larger model parameters \cite{pradeep2023doesscale,cai2025scalewang}, and exploring new training and decoding strategies \cite{sun2023learning, wu2025constrained}.

\header{Limitations}
Although these approaches have achieved competitive performance compared to dense retrieval baselines, several fundamental limitations still exist:
(i) \emph{Mismatch between DocID generation and natural language generation.}
Unlike natural language generation where left-to-right word order encodes rich semantic, tokens of a DocID may not have so strong left-to-right dependencies. It indicates that the auto-regressive left-to-right generation of DocID tokens could limit the retrieval performance. For example, if an incorrect DocID token is generated in early left steps, the following decoding generation steps would become meaningless since the correct DocID would never be generated.
(ii) \emph{Lack of dynamic control over the efficiency–accuracy trade-off.}
Existing auto-regressive based GR methods can only generate DocID tokens one by one during inference, leading to increasingly high computational costs as decoding progresses. 
More importantly, such approaches offer limited flexibility to adapt inference behavior to different operational requirements. In contrast, the ability to dynamically balance retrieval accuracy against inference efficiency is crucial for real-world applications with varied latency and resource constraints.

\header{The Proposed Method}
To address the above challenges, we propose \emph{DiffuGR}, a generative document retrieval framework based on diffusion language models, which formulates DocID generation as a discrete diffusion process.
Specifically, DiffuGR assigns each document a sequence of tokens as its DocID. Two types of DocIDs are considered: learnable DocIDs and text-based DocIDs. During training, DocIDs are corrupted through a stochastic masking process, and a diffusion language model is learned to recover the masked tokens under a retrieval-aware objective function.
At the inference stage, given an input query, DiffuGR generates DocID tokens in parallel and iteratively refines them through a controllable denoising steps, as shown in the bottom part of Figure \ref{fig:ar-vs-dr}. 

DiffuGR offers several appealing properties to GR: 
First, diffusion-based sampling enables parallel generation with iterative refinement: candidate DocID tokens are denoised jointly rather than generated sequentially. This allows the model to revise earlier predictions instead of being irrevocably constrained by early token errors. 
As a result, DiffuGR tends to establish high-confidence DocID tokens early and progressively refine the complete identifier across successive denoising steps.
Second, DiffuGR provides dynamic control over the efficiency–accuracy trade-off by adjusting the number of denoising steps: fewer steps lead to faster inference, while additional steps improve retrieval accuracy.
Finally, the stochastic nature of the denoising process enables DiffuGR to generate multiple plausible DocID candidates, thereby enhancing both the diversity and robustness of retrieval results.

We conduct extensive experiments on NQ320K and MS MARCO datasets. Our key findings are twofold:
(i) \emph{Improved effectiveness performance.}
DiffuGR consistently outperforms strong autoregressive generative retrievers on two widely used benchmark datasets, demonstrating the effectiveness of the proposed approach.
(ii) \emph{Flexible runtime control.}
DiffuGR offers flexible runtime control over the quality-latency trade-off, which enables retrieval systems to perform dynamically scheduling based on load conditions.

\header{Main Contributions}
Our contributions can be summarized as follows: 
\begin{enumerate*}
    \item We introduce diffusion language models as a new paradigm for GR, encouraging non-autoregressive DocID generation. 
    To the best of our knowledge, this is the first work to utilize diffusion language models for end-to-end document retrieval.
    \item We adapt the discrete diffusion paradigm to the generation of DocIDs. By leveraging the non-autoregressive nature and bidirectional context of diffusion models, our method mitigates the structural mismatch issue and alleviates the error accumulation problem inherent in auto-regressive GR.
    \item We conduct extensive empirical comparisons between diffusion-based and auto-regressive GR, validating the effectiveness of DiffuGR.
    \item We demonstrate that DiffuGR provides explicit and flexible control over the quality–latency trade-off, enabling dynamic adaptation of retrieval performance through adjustable denoising budgets.
\end{enumerate*}
\begin{figure*}
    \centering
    \includegraphics[width=0.95\linewidth]{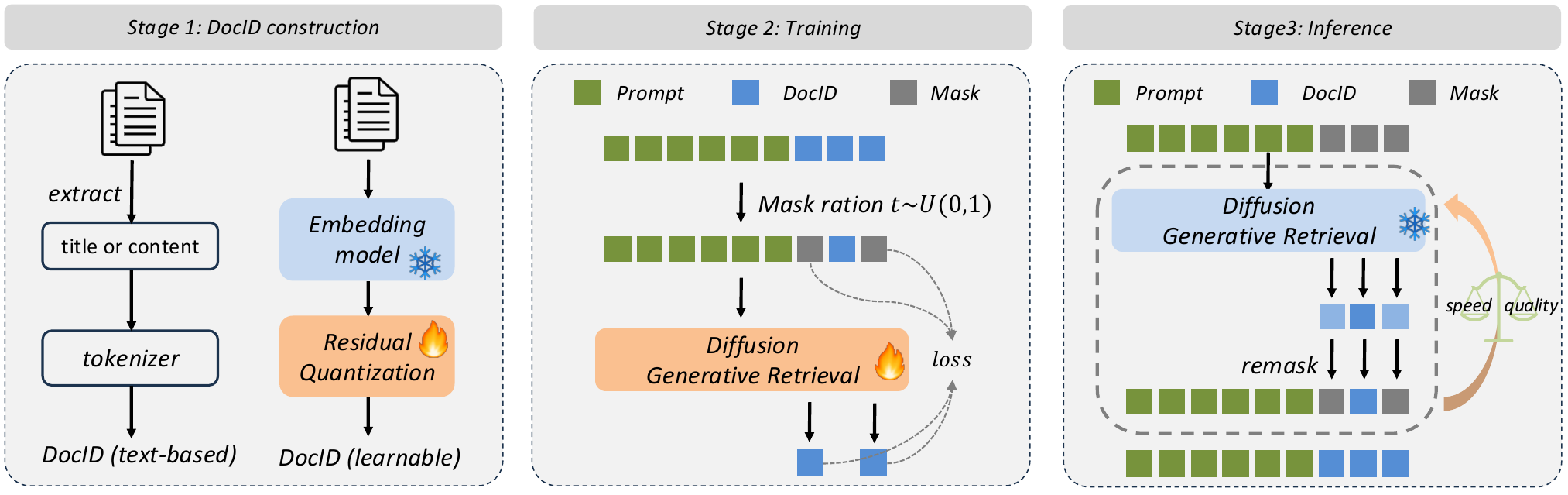}
    \caption{An overview of DiffuGR, consisting DocID construction, model training, and inference. 
    DiffuGR involves two kinds of DocIDs, i.e., text-based DocIDs and learnable DocIDs.
    During training, the model is optimized to recover randomly masked DocID tokens. 
    For inference, DiffuGR  generates DocID tokens in parallel and refines them across multiple denoising steps.}
    \label{fig:pipeline}
\end{figure*}

\section{Related Work}
\subsection{Generative Retrieval}
Generative retrieval (GR) reformulates document retrieval as a generation task, in which a generative model (usually a seq2seq language model) directly generates the identifier of the target document.
Existing GR methods usually use learnable DocIDs, text-based DocIDs, or number-based DocIDs.
Learnable DocIDs are usually constructed upon dense document representations, either through hierarchical $k$-means clustering~\cite{Tay2022TransformerMA, Wang2022ANC}, product quantization~\cite{Zhou2022UltronAU}, residual vector quantization~\cite{TIGER-Rajput, IRGen-Zhang, zeng2023scalable_and_effective}, or progressively learning ~\cite{sun2023learning, yang2023auto_search_indexer}.
This process targets on compressing the semantic information contained in dense document representations into discrete tokens.
Text-based DocIDs use strings, such as title, URL, or keywords, to represent documents.
They naturally carry semantics related to associated documents, and the construction cost is relatively low, without the need for additional manual supervision.
Number-based DocIDs use a single number or a series of numbers to represent documents~\cite{Tay2022TransformerMA, zhou2022dynamicretriever}.
Although methods based on number-based DocIDs offer high retrieval efficiency, particularly when using single number, they are not considered in our experiments due to their poor generalization capability~\cite{zhang2025replication}.
Plenty of work has been conducted based on text-based DocIDs, yielding excellent results\cite{DeCao2020AutoregressiveER, de-cao-etal-2022-multilingual, Bevilacqua2022AutoregressiveSE, Zhou2022UltronAU, Chen2022CorpusBrainPA, zhao2025unifying, UGR-Chen, tang2023inspired_by_learning_strategy, uni_gen, zhang2025replication}.
Despite the success of GR, most existing methods generate DocID tokens in an auto-regressive left-to-right manner. 
Although \citet{qiao2023diffusionret} proposed DiffusionRet which employs the diffusion model to generate a pseudo document for a query, the generation of DocIDs are still auto-regressive.
However, such auto-regressive left-to-right generation of DocID tokens could limit the retrieval performance.
Specifically, if an incorrect DocID token is generated in early left steps, the correct DocID would never be generated.
Based on above considerations, this work explores end-to-end generation of DocIDs through non-autoregressive diffusion language models~\cite{nie2025large, ye2025dream}.

\subsection{Diffusion Models for Text Generation}
Diffusion models are initially introduced in continuous domains such as image generation~\citep{song2020denoising,ho2020denoising}, and their success has inspired extensions to natural language processing by modeling text in continuous embedding spaces~\citep{li2022diffusion,gong2022diffuseq,gong-etal-2023-diffuseq}. 
To better align with the discrete nature of text, discrete diffusion language models were later developed~\citep{austin2021structured,hoogeboom2021argmax,campbell2022continuous}, which progressively corrupt token sequences and then reconstruct the original text during the reverse process. 
Scaling diffusion models to large language models (LLMs) has further extended their applicability. \citet{gulrajani2023likelihoodbased} analyzed scaling laws for continuous diffusion.
~\citet{lou2023discrete} showed that discrete masked diffusion models have achieved perplexities competitive with GPT-2. 
Building on this, large-scale efforts such as DiffuGPT and DiffuLLaMA adopt pretrained  auto-regressive LLMs into diffusion-based frameworks~\citep{gong2025scalingdiffusionlanguagemodels}, while LLaDA~\citep{nie2025large} demonstrated that diffusion models trained from scratch at 8B parameters can rival strong auto-regressive LLMs like LLaMA3-8B. 
While diffusion models have shown promise, effective constrained decoding remains challenging. Recently, \citet{shi2025llada} proposed a specialized beam search for discrete diffusion in generative recommendation. 
In this work, we introduce and validate this decoding paradigm in the context of Generative Retrieval, demonstrating its capability to capture diverse retrieval candidates.

\section{Method}
We begin by formalizing generative document retrieval in Section~\ref{sec:taskform}. Section~\ref{sec:docid} then introduces the construction of DocIDs, followed by a description of the DiffuGR training in Section~\ref{sec:overview}. Finally, we detail the inference process in Section~\ref{sec:inference}. An overview of  DiffuGR is illustrated in Figure~\ref{fig:pipeline}.

\subsection{Task Formulation}
\label{sec:taskform}
Generative retrieval aims to directly generate a document identifier (DocID) given a user query, where the generated DocID uniquely maps a corresponding document.
Formally, let $\bm{D} = \{d_1, d_2, \dots, d_N\}$ denote a document collection, and $\bm{Q} = \{q_1, q_2, \dots, q_M\}$ denote user queries. 
Each document $d \in \bm{D}$ is associated with a unique identifier sequence (DocID), i.e.,  $\bm{z} = \{z_1, z_2, \dots, z_l \}$, where $l$ refers to the length of $\bm{z}$ and each token $z_i \in \bm{z}$ is drawn from residual quantization or token ID of the model vocabulary.
Given a query $q_i \in \bm{Q}$, the model is required to generate the corresponding $\bm{z}$ that identifies the relevant document.

\subsection{Construction of DocIDs}
\label{sec:docid}

A key component of GR is the design of DocIDs.
Unlike dense retrieval methods, GR requires DocIDs to be discrete, compact, and decodable.
To systematically examine the role of DocID design in DiffuGR, we consider two classes of identifiers: \emph{learnable DocIDs} and \emph{text-based DocIDs}.

\subsubsection{Learnable DocIDs}
In this setting, DocIDs are learned through residual quantization over document embeddings.
Let $\bm{d}$ denote the embedding of a document $d$ and $\bm{r}_i$ denote the residual vector of $i$-th step.
The initial residual vector is set as $\bm{r}_0 = \bm{d}$. 
Residual quantization is then performed recursively across multiple levels.

At $i$-th step, we assign $z_i$ as the index number of the  closest code embedding in $i$-th codebook $\bm{E}_i = \{ \bm{e}_i^k \in\mathbb{R}^{h} \mid k = 1, \dots, K_i \}$ to $\bm{r}_{i-1}$, where $K_i$ is the size of $i$-th codebook and $\bm{r}_{i-1}$ denotes the residual vector from the previous step:
\begin{equation}
z_i = \argmin_{k \in \{1,\dots,K_i \}} || \bm{r}_{i-1} - \bm{e}_i^{k} ||^2,
\end{equation}
then the $i$-th residual vector is updated as:
\begin{equation}
\bm{r}_i = \bm{r}_{i-1} - \bm{e}_{i}^{z_i}.
\end{equation}
The original embedding $\bm{d}$ is approximated by adding the code embeddings.
For training of the $i$-th codebook, we adopt the following loss:
\begin{equation}
\bm{L}_{\text{rq}} = \sum_{\bm{D}}\sum_{i=1}^l || \bm{r}_{i-1} - \bm{e}_{i}^{z_i} ||^2.
\label{eq:loss_ml}
\end{equation}

\subsubsection{Text-based DocIDs}
In addition to learnable identifiers, we investigate text-based DocIDs, which are directly constructed from the text of each document.
These DocIDs are lightweight, interpretable, and vocabulary-aligned, making them well-suited for language model generation. 
We consider two representative variants, including:

(1) \emph{Title-based DocIDs.}
The title of each document is directly adopted as the DocID for this document.
Titles are typically concise human-written summaries of the main content, and hence naturally capture the document’s intent. However, not all documents have well-formed or unique titles. To address such issue, DiffuGR incorporates the second kind of text-based DocIDs.

(2) \emph{Leading-token DocIDs.}
The first $n$ tokens of the document content are regarded as the DocID for this document. This alternative exploits the observation that the opening part of a document often introduces its main topic. It should be noted that here we treat such DocIDs as a supplementary method, which is only used to handle samples without document titles.

Here, number-based DocID, e.g, a single number denotes a document, is not used. 
Although it offers high retrieval efficiency, this type of DocID often exhibits poor generalization performance~\cite{zhang2025replication}.

\subsection{Diffusion for Generative Retrieval}
\label{sec:overview}
DiffuGR employs masked diffusion language models to generate DocIDs for a given query. 
It defines a forward process that progressively masks DocID tokens and learns a reverse denoising process that iteratively reconstructs the complete identifier sequence.

\subsubsection{Forward process}
Let $V$ denote the vocabulary size of the masked diffusion language model and $L$ the length of the input sequence.
Given a sequence $\bm{x_0} \in \{0, 1, \dots, V-1\}^L$ and a noise level $t \in [0, 1]$, the forward process randomly and independently masks the tokens, formulated as follows:
\begin{align}
    q_{t|0}(\bm{x_t} | \bm{x_0}) &= \prod_{i=0}^{L-1} q_{t|0}(\bm{x_t^i}|\bm{x_0^i}), \\
    q_{t|0}(\bm{x_t^i} | \bm{x_0^i}) &= 
    \begin{cases}
        1 - t, & \bm{x_t^i} = \bm{x_0^i}, \\
        t, & \bm{x_t^i}= m,
    \end{cases}
\end{align}
where $q_{t|0}(\bm{x_t} | \bm{x_0})$ denotes the transition probability from the original data $\bm{x_0}$ to the noisy data $\bm{x_t}$ at time $t$, $\bm{x^i}$ denotes the $i$-th element of $\bm{x}$, $m$ denotes the mask token~\citep{devlin2018bert} and $q_0(\cdot)$ is the data distribution $p_{\textrm{data}}(\cdot)$.

\subsubsection{Reverse process}
The reverse process iteratively recovers tokens for masked positions, starting from a mask sequence $\bm{x_1}$. 
Let $0\le s < t \le 1$, the reverse process is characterized by
\begin{align}
    q_{s|t}(\bm{x_s}|\bm{x_t}) &= \prod_{i=0}^{L-1} q_{s|t}(\bm{x_s^i}|\bm{x_t}), \\
    q_{s|t}(\bm{x_s^i}|\bm{x_t}) &= 
        \begin{cases}
            1, & \bm{x_t^i} \neq m, \bm{x_s^i} = \bm{x_t^i},\\
            \frac{s}{t}, & \bm{x_t^i} = m, \bm{x_s^i} = m, \\
            \frac{t - s}{t} q_{0|t}(\bm{x_s^i}|\bm{x_t}), & \bm{x_t^i} = m, \bm{x_s^i} \neq m, \\
            0, & \textrm{otherwise}.
        \end{cases}
\label{equ:reverse_process}
\end{align}
Here $q_{0|t}(\cdot|\cdot)$ is the DiffuGR to be learned.
Notably, \citet{ou2024your} revealed an intrinsic property of mask diffusion language model that $q_{0|t}(\cdot|\cdot)$ can be represented by conditional distributions on clean data $p_{\text{data}}(\cdot|\cdot)$ independently from $t$, which means: 
\begin{align}
\label{equ:condition_distribution}
q_{0|t}(\bm{x_0^i}|\bm{x_t}) = p_{\text{data}}(\bm{x_0^i}|\bm{x_t}^{\text{UM}}),
\end{align}
where $\bm{x_t}^{\text{UM}}$ collects all unmasked tokens in noisy data $\bm{x_t}$ and $p_{\text{data}}(\cdot|\cdot)$ is irrelevant to $t$. For example, if $\bm{x_t} =[1, 3, m,2]$, then $\bm{x_t}^{\text{UM}} = [1, 3, \cdot, 2]$ and $p_{\text{data}}(\cdot|[1, 3, \cdot, 2])$ is irrelevant to $t$.

\subsubsection{Training Objective}
The model is trained to reconstruct the original sequence by predicting the masked tokens.
A distribution $p_{\bm{\theta}}(\bm{x_0^i}|\bm{x_t})$ parameterized by $\bm{\theta}$ is employed to approximate $p_{\text{data}}(\bm{x_0^i}|\bm{x_t}^{\text{UM}})$, optimizing the following upper bound on negative log-likelihood:
\begin{equation}
\begin{split}
    - \log 
    & p_{\bm{\theta}}(\bm{x_0}) \leq \\
    & \int_0^1 \frac{1}{t} \mathbb{E}_{q(\bm{x_t}|\bm{x_0})}\left[ \sum_{\{i| \bm{x_t^i} = m\}} - \log p_{\bm{\theta}}(\bm{x_0^i}|\bm{x_t}) \right] d t \triangleq \bm{L}.
\end{split}
\label{equ:loss}
\end{equation}
To put it simply, given a sequence $\bm{x_0}$, we randomly mask a subset of tokens according to a corruption process $q_{t|0}(\bm{x_t} | \bm{x_0})$, where $\bm{x_t}$ denotes the corrupted sequence at time $t$. 
This objective resembles denoising autoencoding, but the iterative refinement aligns naturally with the diffusion process, where the model progressively recovers signal from noise.
As for generative retrieval task, the training data contains query-DocID pairs $(q, \bm{z})$ and document-DocID pairs $(d, \bm{z})$.
We only randomly mask the DocID sequences $\bm{z}$ for training the masked diffusion language model.
Queries are paired with their corresponding document identifiers, and the model learns to generate $\bm{z}$ from $q$. 

\subsection{Model Inference}
\label{sec:inference}

At inference time, the query is concatenated with  fully masked DocID placeholders, and the model attempts to generate DocID tokens in parallel and refine the generation through diffusion-based denoising. 
The total number of denoising iteration is a hyperparameter, which naturally provides diffusion language models with a trade-off between efficiency and quality.
Following \citet{nie2025large, ye2025dream}, we employ uniformly distributed $t$ by default.
Due to the inherent properties of the masked diffusion language model, the generation length $l$ is also a predefined hyperparameter.
In our work, we set $l$ to the maximum length of the DocIDs.

\subsubsection{Diffusion Denoising Strategies}
At the first step, we feed both $q$ and $\bm{z}^1$, where the superscript of $\bm{z}^1$ means that the DocID tokens are all masked. 
Then, for steps from time $t \in (0,1]$ to $s \in [0,t)$, we feed both $q$ and $\bm{z}^{s}$ into the masked diffusion language model.
Here, $\bm{z}^{s}$ denotes the DocID tokens in timestamp $s$, where some tokens remain masked.
The model then predicts all currently masked DocID tokens simultaneously.
Subsequently, DiffuGR re-masks a fraction of the newly predicted tokens, with an expected proportion of ${s}/{t}$. 
This ensures that the transition in the reverse process is consistent with the forward process, enabling accurate sampling.

When it comes to specific implementation of re-masking, a variety of different  strategies can be used.
In this work, DiffuGR explores four denoising scheduling methods that determine which token to be re-masked and regenerated during the reverse diffusion process:
(1) \emph{Random.} 
The random decoding strategy follows a purely random generation order \cite{austin2021structured}, which serves as the default baseline for diffusion-based language models. 
(2) \emph{Maskgit plus.}
This method adopts the confidence-based token scheduling from \cite{chang2022maskgit}, where tokens with the lowest predicted confidence are re-masked.
(3) \emph{Top-$k$ margin.}
This method extends the confidence-based scheduling by ranking tokens according to margin confidence (i.e., the difference between top-1 and top-2 probabilities) \cite{kim2025train}.
(4) \emph{Entropy.}
Tokens with high entropy, indicating low model confidence, are re-masked and re-predicted in later diffusion iterations, whereas low-entropy tokens are preserved to stabilize the denoising process.
In Section~\ref{sec:denoise-strategy}, experiments were conducted to investigate the effects of different denoising strategies on the model performance.

\subsubsection{Pseudo Beam Search for DiffuGR}
\label{sec:query_aug}
While DiffuGR offers efficient parallel token generation, it inherently lacks the fixed left-to-right generation order that enables candidate exploration in beam search. Consequently, DiffuGR cannot directly leverage standard beam search to generate multiple candidate DocIDs. To address this limitation, we explored three strategies to approximate the execution of beam search:
(1) The first strategy leverages query augmentation, where a single query is reformulated into multiple semantically equivalent but lexically diverse variants through a language model.
(2) Candidate Trajectory Extraction (CTE) is the second strategy, which exploits intermediate denoising states within the diffusion process itself. Given a total of $T$ denoising steps, the model produces $T$ intermediate DocID candidates corresponding to different noise levels. By retaining and evaluating these intermediate outputs, DiffuGR can explore multiple decoding hypotheses at negligible additional cost.
(3) The third strategy is an adapted beam search for diffusion model~\citep{shi2025llada}, which has been exploited in generative recommendation. At each iteration, it first identify the position associated with the highest model confidence for generation, followed by conducting beam expansion at this selected position. Through successive rounds of this iterative process, the final recommended top items $B$ are obtained.

The above techniques enable DiffuGR to approximate candidate diversity of beam search without sacrificing its parallel generation property, thus improving retrieval robustness and recall metrics.
However, due to the incompatibility with constrained decoding, None of these methods can guarantee that the generated DocIDs are valid and non-duplicate.
To mitigate these issues, DiffuGR performs validity and deduplication checks on the generated DocIDs.
For each invalid DocID, DiffuGR searches for the nearest valid DocID based on Hamming distance.

\section{Experimental Setup}
\subsection{Datasets and Evaluation Metrics}
We conduct experiments on two well-known document retrieval datasets: NQ320K~\citep{Kwiatkowski2019NaturalQA} and MS MARCO~\citep{Campos2016MSMA}. 

\header{NQ320K}
NQ320K is a widely applied benchmark for evaluating retrieval models~\citep{Tay2022TransformerMA,Wang2022ANC}, based on the Natural Questions (NQ) dataset~\citep{Kwiatkowski2019NaturalQA}. 
NQ320k consists of 320k query-document pairs, where the documents are gathered from Wikipedia pages, and the queries are natural language questions.
Note that our study follows the data processing pipeline as described in DDRO~\citep{mekonnen2025lightweight}, and obtained approximately 100k documents.

\header{MS MARCO}
MS MARCO is a collection of queries and web pages from Bing search.
Similarly to NQ320k, we sample a subset of labeled documents and use their corresponding queries for training. 
We evaluate the models on the queries of the MS MARCO dev set and retrieval on the sampled document subset.

\header{Evaluation Metrics}
On NQ320K and MS MARCO, we use R@\{1,5,10\} and MRR@10 as evaluation metrics, following~\citep{mekonnen2025lightweight}.
R@100 and nDCG@10 are also used in some analytical experiments. 

\subsection{Baselines}
Three types of baselines are considered in our work, including sparse retrieval methods, dense retrieval methods, and generative retrieval methods.

The sparse retrieval baselines are as follows:
\textbf{BM25} uses the tf-idf feature to measure term weights and we use the implementation from \url{http://pyserini.io/}.
\textbf{DocT5Query} expands a document with possible queries predicted by a finetuned T5 with this document as the input.

The dense retrieval baselines are as follows: \textbf{DPR}~\citep{Karpukhin2020DensePR}, a dual-encoder model using the representation of the \code{[CLS]} token of BERT.
\textbf{ANCE}~\citep{Xiong2020ApproximateNN}, an asynchronously updated ANN indexer is utilized to mine hard negatives for training a RoBERTa-based dual-encoder model.
\textbf{Sentence-T5}~\citep{Ni2021SentenceT5SS}, a dual-encoder model that uses T5 to produce continuous sentence embeddings.
\textbf{RepBERT}~\citep{zhan2020repbert}, a BERT-based model that generates fixed-length contextualized embeddings with query-document relevance computed via inner product similarity.

The generative retrieval baselines are as follows:
\textbf{DSI}~\citep{Tay2022TransformerMA}, which represents documents using hierarchical $K$-means clustering results, and indexes documents using the first 32 tokens as pseudo-queries.
\textbf{SEAL}~\citep{Bevilacqua2022AutoregressiveSE} uses arbitrary n-grams in documents as DocIDs, and retrieves documents under the constraint of a pre-built FM-indexer. 
\textbf{DSI-QG}~\citep{Zhuang2022BridgingTG} uses a query generation model to augment the document collection. 
\textbf{Ultron}~\citep{Zhou2022UltronAU} uses a three-stage training pipeline and represents documents through three types of identifiers.
\textbf{ROGER}~\citep{ROGER},which transfers knowledge from a dense retriever to a generative retriever via knowledge distillation.
\textbf{MINDER}~\citep{li2023multiview},which assigns multiple identifiers, including titles, n-grams, and synthetic queries, to documents and pairs them for indexing.
\textbf{LTRGR}~\citep{li2024learning},which trains on pairwise relevance objectives using margin-based ranking loss for optimization.
\textbf{GenRRL}~\citep{zhou2023enhancing},which incorporates pointwise, pairwise, and listwise relevance optimization through reinforcement learning, using document summaries and URLs as DocIDs.
\textbf{DDRO}~\citep{mekonnen2025lightweight}, which enhances generative retrieval by directly aligning generation with document-level relevance estimation.

\subsection{Implementation Details}
\textbf{Models and Hyperparameters.}
In our experiments, we utilize LLaDA~\citep{nie2025large} and Dream~\citep{ye2025dream}  as the diffusion language models.
As for the learnable DocIDs, the length of DocIDs is set as $l=3$ and the size of $i$-th codebook is $K_i=512, i \in \{1,2,3\}$.
For the MS MARCO dataset, the length of learnable DocIDs is set as $l=4$.
Beam size is set to 15.
The maximum number of tokens is $12$ for text-based DocIDs. We optimize the model using AdamW and set the learning rate to $5e-4$.
DeepSeek-V3 is used for the query augmentation.

\header{Data Augmentation}
Following previous work~\citep{Zhuang2022BridgingTG,Wang2022ANC,Wang2021GPLGP}, we use query generation models to generate synthetic (query, document) pairs for training data augmentation.
Following~\citet{sun2023learning}, we use the pre-trained query generation model from DocT5Query~\citep{Cheriton2019FromDT} to augment the NQ320K and MS MARCO datasets. 
In query generation, we use nucleus sampling with parameters $p=0.8, t=0.8$ and generate ten queries for each document in the collection.
\begin{table}[ht]
\centering
\caption{Performance comparison on NQ320K. The best results are shown in bold. The second-best values are underlined. $\dagger$ indicates the result is significantly improved with paired $t$-test at $ p < 0.05 $ level. 
Abbreviations denote DocIDs used for GR baselines: SI -- Semantic ID; PQ -- Product Quantization; NG -- N-grams; TU -- Title + URL. Some results of baselines are copied from ~\cite{mekonnen2025lightweight}, and * denotes that the results are reproduced using public code. R@$k$ is short for Recall@$k$.}
\label{table:nq320k}
\setlength{\tabcolsep}{6pt}
\begin{tabular}{l | cccc}
\toprule
\textbf{Model}  & \textbf{R@1} & \textbf{R@5} & \textbf{R@10} & \textbf{MRR@10} \\
\midrule
\multicolumn{5}{l}{\emph{Sparse retrieval \& Dense retrieval}} \\
\midrule
BM25  & 14.06 & 36.91 & 47.93 & 23.60 \\
DocT5Query & 19.07 & 43.88 & 55.83 & 29.55 \\
DPR & 22.78 & 53.44 & 68.58 & 35.92 \\
ANCE & 24.54 & 54.21 & 69.08 & 36.88 \\
RepBERT & 22.57 & 52.20 & 65.65 & 35.13 \\
Sentence-T5 & 22.51 & 52.00 & 65.12 & 34.95 \\
\midrule
\multicolumn{5}{l}{\emph{Generative retrieval}} \\
\midrule
DSI (SI) & 27.42 & 47.26 & 56.58 & 34.31 \\
DSI-QG (SI) & 30.17 & 53.20 & 66.37 & 38.85 \\
NCI (SI) & 32.69 & 55.82 & 69.20 & 42.84 \\
SEAL (NG) & 29.30 & 54.12 & 68.53 & 40.34 \\
Ultron (TU) & 33.78 & 54.20 & 67.05 & 42.51 \\
Ultron (PQ) & 25.64 & 53.09 & 65.75 & 37.12 \\
ROGER-NCI (SI) & 33.20 & 56.34 &69.80 & 43.45 \\
ROGER-Ultron (TU) & 35.90 & 55.59 &69.86 & 44.92 \\
{MINDER} (SI)  & 31.00 & 55.50 & 65.79 & 43.50 \\
{LTRGR} (SI)  & 32.80 & 56.20 & 68.74 & 44.80 \\
{DDRO} (TU)* & 39.75 &52.07 &54.41 &44.82 \\
{DDRO} (PQ)* & 48.10 &62.58 &66.78 &54.25 \\
\midrule
LLaDA-Learnable & 62.64\rlap{$^\dagger$} &76.05\rlap{$^\dagger$} &77.79\rlap{$^\dagger$} & 66.10\rlap{$^\dagger$} \\

LLaDA-Text &\underline{66.29}\rlap{$^\dagger$} &\underline{ 79.06}\rlap{$^\dagger$} &\underline{81.25}\rlap{$^\dagger$} &\underline{71.86}\rlap{$^\dagger$} \\

Dream-Learnable & 65.63\rlap{$^\dagger$} &78.59\rlap{$^\dagger$}  &79.70\rlap{$^\dagger$} &69.13\rlap{$^\dagger$} \\

Dream-Text & \textbf{69.73}\rlap{$^\dagger$} &\textbf{80.90}\rlap{$^\dagger$} &\textbf{82.82}\rlap{$^\dagger$} &\textbf{74.65}\rlap{$^\dagger$} \\
\bottomrule
\end{tabular}
\end{table}
\begin{table}[t]
\centering
\caption{Performance comparison on MS MARCO. The best results are shown in bold. The second-best values are underlined. $\dagger$ indicates the result is significantly improved with paired $t$-test at $ p < 0.05 $ level.
Abbreviations denote DocIDs used for GR baselines: SI -- Semantic ID; PQ -- Product Quantization; NG -- N-grams; TU -- Title + URL. Some results of baselines are copied from ~\cite{mekonnen2025lightweight}, and * denotes that the results are reproduced using public code.
 R@$k$ is short for Recall@$k$.}
\label{table:msmarco}
\setlength{\tabcolsep}{6pt}
\begin{tabular}{l | cccc}
\toprule
\textbf{Model}  & \textbf{R@1} & \textbf{R@5} & \textbf{R@10} & \textbf{MRR@10} \\
\midrule
\multicolumn{5}{l}{\emph{Sparse retrieval \& Dense retrieval}} \\
\midrule
BM25  & 18.94 & 42.82 & 55.07 & 29.24 \\
DocT5Query & 23.27 & 49.38 & 63.61 & 34.81 \\
DPR & 29.08 & 62.75 & 73.13 & 43.41 \\
ANCE & 29.65 & 63.43 &74.28 & 44.09 \\
RepBERT & 25.25 & 58.41 & 69.18 & 38.48 \\
Sentence-T5 & 27.27 & 58.91 & 72.15 & 40.69 \\
\midrule
\multicolumn{5}{l}{\emph{Generative retrieval}} \\
\midrule
DSI (SI) & 25.74 & 43.58 & 53.84 & 33.92 \\
DSI-QG (SI) & 28.82 & 50.74 & 62.26 & 38.45 \\
NCI (SI) & 29.54 & 57.99 & 67.28 & 40.46 \\
SEAL (NG) & 27.58 & 52.47 & 61.01 & 37.68 \\
Ultron (TU) & 29.82 & 60.39 & 68.31 & 42.53 \\
Ultron (PQ) & 31.55 & 63.98 & 73.14 & 45.35 \\
ROGER-NCI (SI) & 30.61 & 59.02 & 68.78 & 42.02 \\
ROGER-Ultron (TU) & 33.07 & 63.93 &\textbf{75.13} & 46.35 \\
{MINDER} (SI)  & 29.98 & 58.37 & 71.92 & 42.51 \\
{LTRGR} (SI)  & 32.69 &64.37 & 72.43 &47.85 \\
{DDRO} (PQ)* & 31.06 &62.87 &71.91 &44.31 \\
{DDRO} (TU)* & 37.76 &65.83 &73.95 &49.32 \\
\midrule
LLaDA-Learnable
&43.42\rlap{$^\dagger$} &63.79 &65.20 &52.29\rlap{$^\dagger$} \\

LLaDA-Text
&\underline{45.72}\rlap{$^\dagger$} &\underline{66.60}\rlap{$^\dagger$} &71.35 &53.38\rlap{$^\dagger$} \\

Dream-Learnable
&\textbf{46.83}\rlap{$^\dagger$} &63.48 &64.49 &\underline{55.15}\rlap{$^\dagger$} \\ 

Dream-Text
&45.38\rlap{$^\dagger$} &\textbf{69.58}\rlap{$^\dagger$} &\underline{74.55} &\textbf{55.60}\rlap{$^\dagger$}\\

\bottomrule
\end{tabular}
\end{table}

\section{Results and Analysis}
In this section, we present the experimental results to systematically evaluate the performance of DiffuGR and analyze the contribution of its core components. We list the following research questions to guide our experiments: 
\begin{enumerate}[label=\textbf{RQ\arabic*},nosep,leftmargin=*]
   \item How does DiffuGR perform compared to baseline methods?\label{Q:main-results}
   \item How does DiffuGR balance the quality–latency tradeoff?
   \label{Q:speed-quality}
   \item How does DiffuGR generate top-$k$ results with pseudo beam search?\label{Q:pseudo-beam-search}
   \item How do hyperparameters affect DiffuGR performance, including decoding strategies and model scales?\label{Q:hyper-parameter}
\end{enumerate}
\noindent

\subsection{Performance Comparison (RQ1)}

\header{Results on NQ320K and MS MARCO}
Table~\ref{table:nq320k} and Table~\ref{table:msmarco} summarize the performances comparison on the NQ320K and  MS MARCO, respectively. 
We draw the following observations:
(1) The proposed DiffuGR significantly outperforms all generative retrieval baselines on two datasets.
Specially, DiffuGR achieving 21.63 and 20.4 higher R@1 and MRR@10, respectively, compared to the best-performing baseline DDRO~\cite{mekonnen2025lightweight} on NQ320K dataset.
Even on the challenging MS MARCO dataset, DiffuGR also achieves favorable performance.
This highlights the effectiveness of discrete diffusion paradigm in generative document retrieval task.
(2) Different types of DocID affect the performance of DiffuGR. 
Overall, using text DocIDs yields better performance than using learnable DocIDs.
This is likely because text-based DocIDs can directly leverage the semantic space of existing models, whereas learnable DocIDs require vocabulary expansion and semantic remapping, which further processes learning difficulty.
Furthermore, we notice that different types of DocIDs exert a considerable impact on the performance of DiffuGR on the NQ320K dataset. We hypothesize that this may be attributed to the fact that the documents in the NQ320K dataset are equipped with well-formulated titles, which generally serve as a highly concise summary of the documents; thus, text-based DocIDs can yield better performance.
Note that the focus of this paper lies in non-autoregressive generation of DocIDs other than the construction.
(3) In addition, we also observe that most methods perform worse on the MS MARCO dataset than on NQ320K. 
This may be attributed to the fact that the MS MARCO dataset is a real web search dataset.
For some reason, the titles of some documents may not summarize the document content well, and some documents even do not have a title at all. Furthermore, the queries provided by users also exhibit a certain degree of ambiguity and diversity, which impairs the performance.

\begin{table}[ht]
\centering
\setlength\tabcolsep{10pt}
\caption{Zero-shot evaluation for BEIR dataset.
The best results of generative retrieval are shown in bold.
the second-best values of generative retrieval are underlined.
Abbreviations denote DocIDs used for GR baselines: Naive - Naive String; Atomic - Atomic ID; SI - Semantic ID; PQ - Product Quantization; TU - Title + URL.  
* denotes that the results are reproduced using public code.}
\label{table:zero-shot-beir}
\begin{tabular}{l | c | cc }
\toprule
& \multicolumn{3}{c}{\textbf{BEIR (nDCG@10)}}  \\

\textbf{Method} &\textbf{Average} &\textbf{Arg} &\textbf{SciDocs} \\

\midrule
\multicolumn{4}{l}{\emph{Sparse retrieval}} \\
\midrule
BM25 &21.95 &29.1 & 14.8 \\
DocT5Query &25.55 &34.9 & 16.2 \\

\midrule
\multicolumn{4}{l}{\emph{Generative retrieval}} \\
\midrule
GENRE &0.30 &0.0  &0.6 \\
DSI (Naive)* &0.10 &0.1  &0.1 \\
DSI (Atomic)* &0.15 &0.2 &0.1 \\
DSI (SI)* &3.85 &1.8  &\underline{5.9} \\
NCI (SI)*  &1.05 &0.9  &1.2 \\
DDRO (PQ)* &0.03 &0.04  &0.02 \\
DDRO (TU)* &0.55 &1.0  &0.1 \\

\midrule
LLaDA+Learnable &\textbf{10.35} &\textbf{13.8} &\textbf{6.9} \\
LLaDA+Text &\underline{9.25} &\underline{13.1} &5.4 \\
Dream+Learnable &7.10 &10.4 &3.8 \\ 
Dream+Text &4.10 &6.2 &2.0 \\
\bottomrule
\end{tabular}
\end{table}
\header{Zero-shot performance on BEIR}
Table~\ref{table:zero-shot-beir} shows the zero-shot retrieval performance on the BEIR~\cite{thakur2021beir} dataset after training on the NQ320K dataset.
On the BEIR benchmark, DiffuGR consistently achieves competitive performance compared to other generative retrieval methods and attains the best average results. 
Its iterative denoising process generates high-confidence tokens step by step, which effectively reduces error accumulation and leads to better zero-shot generalization. In contrast, auto-regressive generative retrieval methods are more susceptible to cascading errors, resulting in inferior zero-shot performance.
We further observe that generative retrieval methods based on semantic IDs exhibit relatively better zero-shot performance.
We hypothesize that this may be due to the short length of semantic IDs, which reduces the retrieval difficulty of new documents.
However, sparse retrieval methods generally outperform generative retrieval models in the zero-shot setting, highlighting generalization as a key challenge of generative retrieval.

\begin{table}[ht]
\centering
\setlength\tabcolsep{5pt}
\caption{Evaluation of catastrophic forgetting without beam search.
The best results are shown in bold.
The second-best values are underlined.
Abbreviations denote DocIDs
used for GR baselines: SI – Semantic ID; PQ – Product Quanti-
zation; TU – Title + URL.
$\dagger$ indicates the result is significantly improved with paired $t$-test at $ p < 0.05 $ level.
* denotes that the results are reproduced using public code.
}
\label{table:zero-shot-nq-forget-main}
\begin{tabular}{l | cccc @{}}
\toprule
& \multicolumn{4}{c}{\textbf{$\text{NQ320K}_{old}$}}  \\

\textbf{Method} & \textbf{R@1} & \textbf{R@10} &\textbf{MRR@10} &\textbf{nDCG@10} \\
\midrule
\multicolumn{4}{l}{\emph{Training on $\text{NQ320K}_{old}$}} \\
\midrule
DSI (SI)* &29.31 &33.58 &32.96 &32.87 \\
DDRO (TU)* &43.82 &44.35 &44.17 &44.21 \\
DDRO (PQ)* &49.97 &51.53 &51.32 &51.28 \\
LLaDA+Learnable &66.02 &69.01 &67.44 &67.85 \\
LLaDA+Text &\underline{69.32} &\underline{69.59} &\underline{69.43} &\underline{69.47} \\ 
Dream+Learnable &65.80 &68.76 &67.20 &67.60 \\
Dream+Text &\textbf{69.57}\rlap{$^\dagger$} &\textbf{69.77}\rlap{$^\dagger$} &\textbf{69.65}\rlap{$^\dagger$} &\textbf{69.65}\rlap{$^\dagger$} \\
\midrule
\multicolumn{4}{l}{\emph{Continual Training on $\text{NQ320K}_{new}$}} \\
\midrule
DSI (SI)* &11.53 &13.98 &13.83 &13.23 \\
DDRO (TU)* &14.98 &15.72 &15.64 &15.71 \\
DDRO (PQ)* &16.45 &17.91 &17.57 &17.60 \\
LLaDA+Learnable &19.95 &20.87 &20.38 &20.51 \\
LLaDA+Text &\underline{39.72} &\underline{39.87} &\underline{39.78} &\underline{39.81} \\
Dream+Learnable &22.06 &23.15 &22.58 &22.73 \\
Dream+Text &\textbf{45.12}\rlap{$^\dagger$} &\textbf{45.27}\rlap{$^\dagger$} &\textbf{45.18}\rlap{$^\dagger$} &\textbf{45.21}\rlap{$^\dagger$} \\

\bottomrule
\end{tabular}
\end{table}
\header{Catastrophic forgetting problem}
To investigate whether catastrophic forgetting occurs when applying continual learning to adapt DiffuGR to new documents, we randomly select 10000 documents from NQ320K to construct a new dataset $\text{NQ320K}_{new}$.
The rest of the documents are then labeled $\text{NQ320K}_{old}$.
As shown in Table~\ref{table:zero-shot-nq-forget-main},  after continual training on the $\text{NQ320K}_{new}$, the performance of generative retrieval methods on the $\text{NQ320K}_{old}$ significantly declined. 
It suggests that these methods encounter a distinct catastrophic forgetting issue.
This suggests that merely conducting continuous training on new documents for generative retrieval does not represent an optimal approach.
This is a common challenge in the field of generative retrieval.
The approach to tackle this challenge lies outside the purview of this paper. We thus designate it as future work.
However, DiffuGR still maintains competitive performance. We hypothesize that bidirectional attention and non-autoregressive generation mitigate the impact of catastrophic forgetting on DiffuGR.
Furthermore, we observe that the DiffuGR with text-based DocID demonstrates superior anti-catastrophic forgetting performance compared with the learnable DocID counterpart, as evidenced by the reduced magnitude of performance decline of DiffuGR following continuous learning.
We hypothesize that this may be attributed to the higher lexical diversity of text-based DocIDs, which reduces the risk of being overwritten during continuous learning.

\begin{figure}
    \centering
    \includegraphics[width=0.9\linewidth]{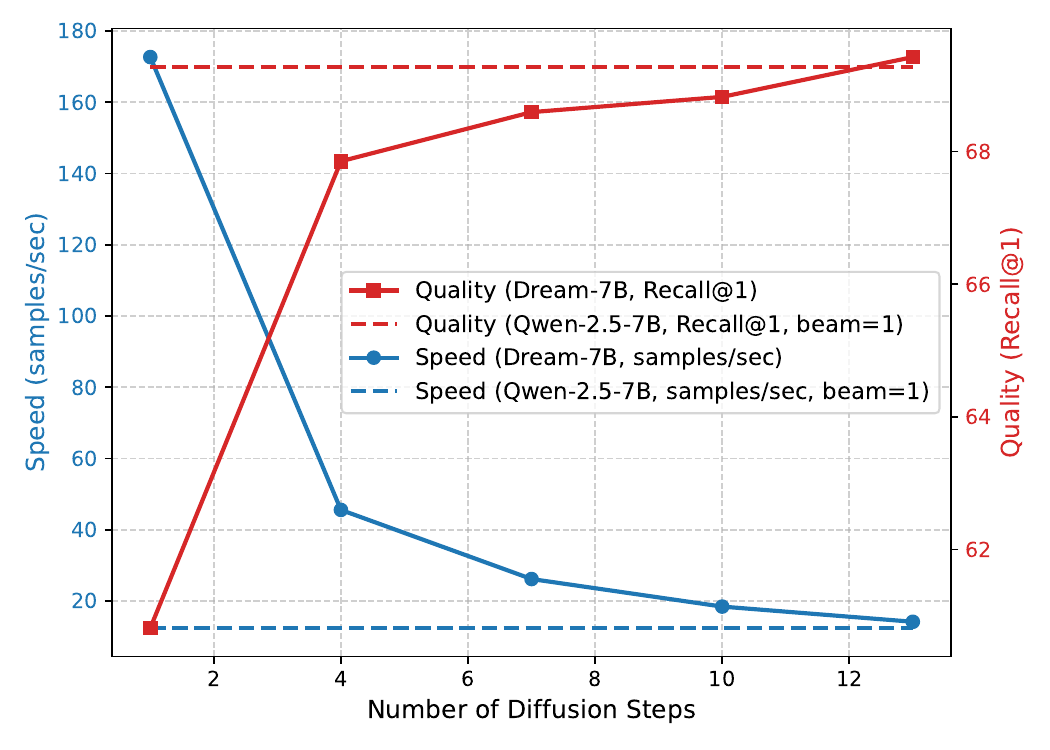}
    \caption{Quality-speed comparison on NQ320K for Dream 7B and Qwen2.5-7B. By adjusting the denoising steps, the performance of DiffuGR can be flexibly tuned towards either speed or quality.}
    \label{fig:tradeoff}
\end{figure}

\begin{figure}
    \centering
    \includegraphics[width=0.9\linewidth]{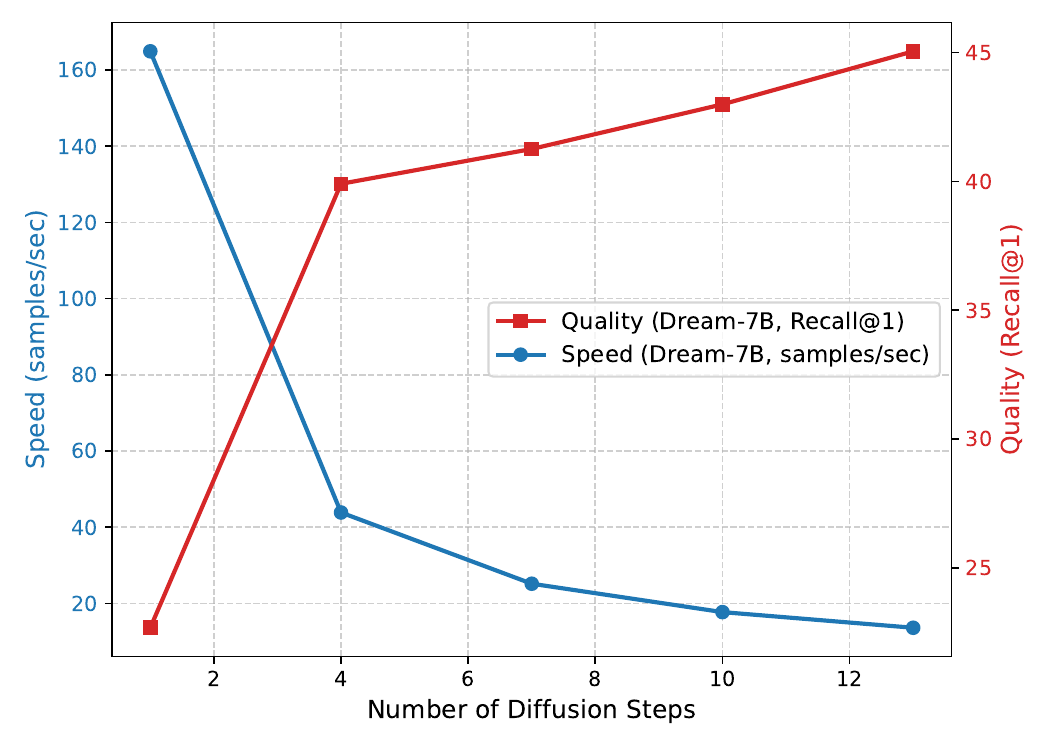}
    \caption{Quality-speed tradeoff on MS MARCO.
    By adjusting the denoising steps, the performance of DiffuGR can be flexibly tuned towards either speed or quality.}
    \label{fig:tradeoff-msmarco}
\end{figure}

\subsection{Quality-Speed Tradeoff (RQ2)}
To answer \ref{Q:speed-quality},
we conducted an ablation study to evaluate the speed and quality with varying numbers of denoising iterations. 
Figure~\ref{fig:tradeoff} and Figure ~\ref{fig:tradeoff-msmarco} show the results on NQ320K and MS MARCO, respectively.
As shown in Figure~\ref{fig:tradeoff} and Figure~\ref{fig:tradeoff-msmarco}, increasing the number of iterations consistently improves DiffuGR performance, with higher R@1 scores achieved at larger iteration steps. 
However, the computational cost grows with the increase of iterations, leading to longer inference latency. 
For example, reducing the steps by half results in a noticeable decrease in latency while maintaining competitive retrieval accuracy, whereas further reducing the steps yields faster decoding at the expense of quality degradation.
For comparison, we tested Qwen-2.5-7B R@1 and the number of samples processed per second on NQ320k.
It can be observed that although Qwen achieved relatively competitive performance, its speed is relatively slow and it fails to flexibly balance the tradeoff between performance and speed.
These findings underscore a crucial advantage of DiffuGR: its controllable generation process enables the dynamic adjustment of inference cost, rendering it appropriate for scenarios characterized by efficiency requirements.

\begin{table}[ht]
\centering
\setlength\tabcolsep{6pt}
\caption{Effect of pseudo beam search on NQ320K. 
CTE is short for candidate trajectory extraction. BS is short for beam search.
The best results are shown in bold. $\dagger$ indicates the result is significantly improved with paired $t$-test at $ p < 0.05 $ level. R@$k$ is short for Recall@$k$.}
\label{table:beam-search}

\begin{tabular}{l | cccc }

\toprule
\textbf{Method} &\textbf{R@1} &\textbf{R@10} &\textbf{R@100} &\textbf{MRR@10} \\

\midrule
LLaDA-Text &66.15 &66.90  &66.90 &66.26 \\
+ query aug &\textbf{66.62} &66.96  &66.96 &66.33 \\

+ CTE  &66.57 &67.34  &67.34 &66.97 \\

+ diffusion BS &66.29 &\textbf{81.25}\rlap{$^\dagger$} &\textbf{84.45}\rlap{$^\dagger$} &\textbf{71.86}\rlap{$^\dagger$} \\

\midrule

Dream-Text &69.47 &69.78  &69.78 &69.57\\
+ query aug &69.50 &69.82  &69.82 &69.71 \\
+ CTE &69.57 &69.83 &69.83 &69.72 \\
+ diffusion BS &\textbf{69.73} &\textbf{82.82}\rlap{$^\dagger$} &\textbf{85.83}\rlap{$^\dagger$} &\textbf{74.65}\rlap{$^\dagger$}\\

\bottomrule
\end{tabular}
\end{table}

\begin{table}[ht]
\centering
\caption{Effect of pseudo beam search on MS MARCO.
CTE is short for candidate trajectory extraction. BS is short for beam search.
The best results are shown in bold. $\dagger$ indicates the result is significantly improved with paired $t$-test at $ p < 0.05 $ level. R@$k$ is short for Recall@$k$. }
\label{table:beam-search-msmarco}

\begin{tabular}{l | cccc }

\toprule
Method &R@1 &R@10 & R@100 & MRR@10 \\

\midrule
LLaDA-Learnable &42.97 &43.60  &43.60 &43.24 \\
+ query aug &43.21 &43.68 &43.71 &43.26 \\
+ CTE &\textbf{43.59} &43.89 &43.89 &43.27 \\
+ diffusion BS &43.42 &\textbf{65.20}\rlap{$^\dagger$} &\textbf{65.68}\rlap{$^\dagger$} &\textbf{52.29}\rlap{$^\dagger$} \\
\midrule

Dream-Learnable &46.60 &46.92  &46.92 &45.96\\
+ query aug &46.89 &46.95 &46.98 &45.97\\

+ CTE &46.62 &46.93 &46.92 &45.97 \\
+ diffusion BS &\textbf{46.83} &\textbf{64.49}\rlap{$^\dagger$} &\textbf{64.70}\rlap{$^\dagger$} &\textbf{55.15}\rlap{$^\dagger$} \\

\bottomrule
\end{tabular}
\end{table}

\subsection{Effect of Pseudo Beam Search (RQ3)}\label{sec:beam-search}
To answer \ref{Q:pseudo-beam-search} and assess the effectiveness of the proposed pseudo beam search strategies,
experiments are conducted on DiffuGR with the three beam search strategies.
Table~\ref{table:beam-search} and Table~\ref{table:beam-search-msmarco} report the Recall@\{1,10,100\} and MRR@10 on NQ320K and MS MARCO, respectively.
The results indicate that three strategies enhance retrieval accuracy compared to the vanilla baseline. However, distinct methods lead to significantly different degrees of performance improvement:
(1) Query augmentation method yields marginal gains by introducing diversity in query representations. 
We find the reason lies in the fact that most of the DocIDs generated by this method are duplicates. 
Duplicate DocIDs result from the trait of high-performance generative retrieval models: they produce identical DocIDs for semantically identical queries with different expressions or wordings.
We hypothesize that this method could effectively enhance model performance when dealing with ambiguous queries. This is because query augmentation can mitigate the ambiguity in queries, thereby improving the model's top‑k performance. However, research on ambiguous queries falls beyond the scope of this work, and we defer it to future studies.
(2) The Candidate Trajectory Extraction (CTE) method enhances the top‑k performance of DiffuGR by leveraging intermediate results from the denoising process. However, its improvements on the R@10/100 and MRR@10 metrics remain limited. 
This limitation stems from two main factors. 
First, DocIDs produced in denoising are noisy.
Second, Once a correct DocID is generated early in decoding, the model tends to retain it in later steps, causing substantial repetition of DocIDs.
A key advantage of the method is that it introduces no additional computational overhead during inference.
Since it only requires retaining the outputs from the forward denoising process, it remains applicable even under resource‑constrained inference scenarios.
(3) The Diffusion Beam Search method achieves significant performance improvement on the R@10/100 and MRR@10 metrics.
It is important to note, however, that this method requires substantial computational resources, primarily for two reasons.
First, it expands the number of inference samples by a factor of B, thereby increasing computational costs by approximately B times.
Second, the selection process for each token involves applying a top-k algorithm, which introduces additional computational overhead.

\begin{table}[!th]
\centering
\setlength\tabcolsep{10pt}
\caption{Impact of denoising strategies on NQ320K without beam search. The best results are shown in bold. $\dagger$ indicates the result is significantly improved with paired $t$-test at $ p < 0.05 $ level. $\ddagger$ denotes the default setting. R@$k$ is short for Recall@$k$.}
\label{table:denoise-st}

\begin{tabular}{l | ccc }

\toprule
\textbf{Method} &\textbf{R@1} &\textbf{R@10} &\textbf{R@100} \\

\midrule
\multicolumn{4}{l}{Dream-Text} \\
+ random & 67.81 &68.12  &68.12 \\
+ maskgit plus\rlap{$^\ddagger$} & \textbf{69.47}\rlap{$^\dagger$} &\textbf{69.78}\rlap{$^\dagger$}  &\textbf{69.78}\rlap{$^\dagger$} \\

+ topk margin &69.38 &69.69  &69.69 \\

+ entropy &69.02 &69.33 &69.33 \\

\bottomrule
\end{tabular}
\end{table}

\begin{table}[!t]
\centering
\setlength\tabcolsep{10pt}
\caption{Impact of denoising strategies on MS MARCO without beam search. The best results are shown in bold. $\dagger$ indicates the result is significantly improved with paired $t$-test at $ p < 0.05 $ level. $\ddagger$ denotes the default setting. R@$k$ is short for Recall@$k$.}
\label{table:denoise-st-msmarco}

\begin{tabular}{l | ccc }

\toprule
Method & R@1 & R@10 & R@100 \\

\midrule
\multicolumn{4}{l}{Dream-Text} \\
+ random & 39.77 &41.76  &42.04 \\

+ maskgit plus\rlap{$^\ddagger$} & \textbf{45.05}\rlap{$^\dagger$} &47.13  &47.21 \\

+ topk margin &44.97 &\textbf{47.21}\rlap{$^\dagger$}  &\textbf{47.52}\rlap{$^\dagger$} \\

+ entropy &44.09 &46.25 &46.56 \\

\bottomrule
\end{tabular}
\end{table}

\subsection{Impact of Hyperparameters (RQ4)}
\subsubsection{Impact of Denoising Strategies}\label{sec:denoise-strategy}
Table~\ref{table:denoise-st} and Table \ref{table:denoise-st-msmarco} show the impact of denoising strategies on NQ320K and MS MARCO, respectively. 
As shown in Table~\ref{table:denoise-st} and Table~\ref{table:denoise-st-msmarco}, different denoising strategies yield distinct retrieval performance under the same testing configuration. 
We observe that the \textit{maskgit plus} and \textit{topk margin} strategies consistently outperform the random baseline, demonstrating the benefit of confidence-guided denoising in reducing uncertainty accumulation during generation. 
The \textit{entropy} strategy achieves comparable performance, suggesting that both local certainty (margin) and global uncertainty (entropy) are effective cues for progressive denoising. 
In practical applications, these strategies can be adopted to provide users with two types of query styles: precision-oriented and diversity-oriented.

\begin{figure}
    \centering
    \includegraphics[width=0.9\linewidth]{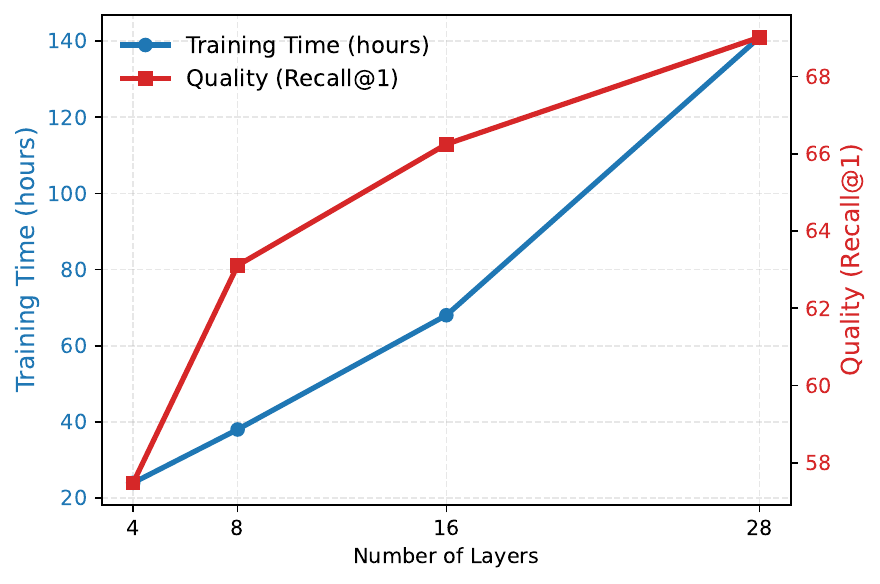}
    \caption{Retrieval performance of DiffuGR on NQ320K under different model scales. Larger models consistently yield higher retrieval accuracy, demonstrating that increased model sizes enhance the retrieval process. However, larger models also incur substantially higher training costs, revealing a clear tradeoff between effectiveness and efficiency.}
    \label{fig:model-size}
\end{figure}
\subsubsection{Impact of Model Scales}
In order to explore the impact of model scales on the performance of DiffuGR, a series of experiments were carried out, in which diffusion language models with different parameter scales were utilized.
Specifically, we experiment with Dream  of different model depths, comprising 4, 8, and 16 transformer layers, derived from the original 28-layer Dream architecture.
Figure~\ref{fig:model-size} presents the results on NQ320K.
We can observe that larger models consistently outperform smaller ones on R@1, demonstrating that increasing model scales enhances the diffusion-based retrieval process.
However, we also observe a significant increase in training time as model size grows, which means training consumes more costs.

\begin{table}[!t]
\centering
\setlength\tabcolsep{8pt}
\caption{Perform comparison between diffusion-based generation and autoregressive generation on the NQ320K dataset without beam search.
R@$k$ is short for Recall@$k$. Best results are shown in bold.
}
\label{table:dlm-vs-arlm}
\begin{tabular}{l | cccc }

\toprule
\textbf{Method} &\textbf{R@1} &\textbf{R@10} &\textbf{R@100} \\

\midrule
Qwen2.5-7B-Text &68.96 &69.27 &69.27 \\

Dream-Learnable & 65.63 &68.54  &68.54 \\

Dream-Text & \textbf{69.73} & \textbf{69.78} & \textbf{69.78} \\

\bottomrule
\end{tabular}
\end{table}

Besides, we also conducted experiments to 
assess the effectiveness of diffusion language models compared to auto-regressive large language models under the same model size.
Specifically, we benchmark our diffusion-based model against Qwen2.5-7B.
This setup ensures that performance differences can be attributed to the generation paradigms rather than model scales.
As shown in Table~\ref{table:dlm-vs-arlm}, Dream+Text achieves better R@1 compared to that of Qwen2.5-7B.
The results indicate that, within the same parameter scale, DiffuGR is comparable with auto-regressive models. Moreover, it offers a novel research direction for generative retrieval.

\section{Conclusion}

In this paper, we have proposed a generative document retrieval framework based on diffusion language models, termed DiffuGR.
DiffuGR is compatible with two different types of DocIDs: learnable DocID and text-based DocID.
DiffuGR employs masked diffusion language models to generate
DocIDs for a given query. It provides a forward process that progres-
sively masks DocID tokens and learns a reverse denoising process
that iteratively reconstructs the complete identifier sequence.
We have conducted extensive experiments to examine the performance of DiffuGR on two widely used benchmarks.
Our results show that DiffuGR achieves outstanding performance compared with strong auto-regressive baselines, while its generation paradigm is more suitable for generative document retrieval tasks. 
Moreover, the generation quality and latency of DocIDs can be flexibly controlled by adjusting the number of denoising iterations. 

The limitations of this work are threefold.
First, we do not optimize DocID construction, as our primary focus is on non-autoregressive DocID generation.
Second, DiffuGR currently does not support constrained decoding, which may result in the generation of invalid DocIDs.
Third, DiffuGR still suffers from catastrophic forgetting under continual training, which is also shared by existing generative retrieval methods.
Addressing these limitations constitutes important directions for future work. In particular, we plan to explore diffusion-aware DocID construction strategies and validity-preserving decoding mechanisms, as well as improve the robustness of diffusion-based generative retrievers under dynamic and continually evolving corpora. We hope this work will provide useful insights and serve as a foundation for future research in generative retrieval.

\bibliographystyle{ACM-Reference-Format}
\bibliography{references}

\end{document}